\documentclass{optica-article}

\journal{opticajournal} 

\articletype{Research Article}
\usepackage{fancyhdr}
\usepackage{lineno}
\usepackage{amsmath}
\usepackage{bm}
\usepackage{tikz,xcolor,hyperref}
\usepackage{hyperref}
\usepackage{lastpage}
\definecolor{lime}{HTML}{A6CE39}
\DeclareRobustCommand{\orcidicon}{%
	\begin{tikzpicture}
	\draw[lime, fill=lime] (0,0) 
	circle [radius=0.16] 
	node[white] {{\fontfamily{qag}\selectfont \tiny ID}};
	\draw[white, fill=white] (-0.0625,0.095) 
	circle [radius=0.007];
	\end{tikzpicture}
	\hspace{-2mm}
}

\foreach \x in {A, ..., Z}{%
	\expandafter\xdef\csname orcid\x\endcsname{\noexpand\href{https://orcid.org/\csname orcidauthor\x\endcsname}{\noexpand\orcidicon}}
}

\begin{document}
\pagestyle{fancy}
\fancyhead{} 
\fancyhead[RO,LE]{\textbf{Opt. Express 32, 45786-45800 (2024)  \href{https://doi.org/10.1364/OE.541271}{DOI: 10.1364/OE.541271}}}
\fancyfoot{} 
\fancyfoot[LE,LO]{\href{https://doi.org/10.1364/OE.541271}{DOI: 10.1364/OE.541271}}
\fancyfoot[CO,CE]{\thepage \vspace{2em} of \pageref{LastPage}}
\fancyfoot[RO,RE]{\href{https://sajid.buet.ac.bd}{sajid.buet.ac.bd}}
\title{Lithium Niobate Photonic Topological Insulator-based Multi-Wavelength Optical Demultiplexer with Piezoelectric Switch-Off}

\author{Prithu Mahmud,\authormark{1,2,\dag} \orcidA{} Kaniz Fatema Supti,\authormark{1,2,\dag} \orcidB{} and \\ Sajid Muhaimin Choudhury\authormark{1,*} \orcidC{}}

\address{\authormark{1}Department of Electrical and Electronic Engineering, Bangladesh University of Engineering and Technology,  Dhaka 1205, Bangladesh\\
\authormark{2}Department of Computer Science and Engineering, BRAC University, Dhaka 1212, Bangladesh\\
}
\authormark{\dag}The authors contributed equally to this work.\\
\email{\authormark{*}sajid@eee.buet.ac.bd} 

\medskip

\begin{abstract*} 
Photonic topological insulators provide unidirectional, robust, wavelength-selective transport of light at an interface while keeping it insulated at the bulk of the material. The non-trivial topology results in an immunity to backscattering, sharp turns, and fabrication defects. This work leverages these unique properties to design a 2-channel optical demultiplexer based on a lithium niobate photonic topological insulator with piezoelectric switch-off capabilities. A photonic topological insulator design for the demultiplexer allows for good wavelength selectivity, crosstalk as low as $-$54 dB, and better isolation between output channels. The primary operating wavelengths presented are the telecommunication wavelengths of 1310 nm and 1550 nm, but the use of the lithium niobate material allows operation at multiple operating wavelengths. Furthermore, we propose a post-fabrication method to switch off the topological protection and, thus, optical transmittance via an applied voltage utilizing the inverse piezoelectric effect of lithium niobate. This work will contribute to advancing lithium niobate integrated photonics and developing efficient, multi-wavelength, electrically controlled optical communication systems and integrated photonic circuits.

\end{abstract*}

\section{Introduction}
Manipulation of electromagnetic waves is vital for modern state-of-the-art technologies. Photonic topological insulators (PTI) \cite{Khanikaev:2013, Ozawa_2019, Cheng:2016,9445781, rechtsman2013photonic, Yang, khanikaev2017two,lu2014topological,fu2011topological, Gal, Ni:21, Fu2, Lustig} are a unique class of materials with exciting properties that are already contributing towards this objective. PTIs are inspired by their electronic counterpart in condensed matter physics \cite{Moore:2010, Kane:2005, Tokura:2019}. The field of topological insulators (TI) began after the discovery of the quantum Hall effect (QHE) in a 2D electron gas when subjected to periodic potentials and external magnetic fields \cite{Thouless, Haldane}. QHE shows that conductivity is fundamentally discrete \cite{Hasan}, and this discrete nature of conductivity is highly robust to deformations in the bulk of the material. If two materials with unequal topological invariants are placed together, a highly robust mode can be seen at the edge of the materials \cite{Hasan}. These "edge" modes exist inside the band gap of the non-trivial materials and arise due to the sudden change of the topological invariant across the boundary. This confines electrons to a robust unidirectional flow within these boundaries \cite{Bansil}. The topological behavior is due to the wave nature of electrons \cite{Wang}, and hence electromagnetic systems with analogous topological properties can be formed \cite{wang2009observation}. PTIs are photonic systems with nonzero topological invariants that exhibit robust edge modes in their band gap. The topological invariant is known as the Chern number. Compared to traditional photonic waveguides with zero Chern numbers, the waveguide formed at the interface between two materials with unequal Chern numbers will be immune to backscattering and defects. This is due to the topological phase transition that must happen at the interface of these materials \cite{Ozawa_2019}.
The flow of light at these interfaces is also immune to sharp turns because the edge modes are tied to the bulk Chern numbers and are, hence, independent of the shape of the interface. The design proposed in this paper is that of a valley PTI. Traditional Chern PTIs are the direct analogs of QHE \cite{Haldane2008} while valley PTIs \cite{Noh:2018, Ma_2016} are analogs of the quantum valley Hall effect studied in graphene-like materials. A controllable band gap can be achieved by breaking inversion symmetry in a graphene-like photonic crystal. Such a structure exhibits opposite Berry curvature at the $K$ and $K'$ points \cite{Blanco}. This allows selective coupling to the $K$ or $K'$ valleys, resulting in a unidirectional edge mode. Two complementary valley PTIs with opposite valley Chern numbers can form a robust edge mode at their interface. 

Lithium niobate (LN) is a popular synthetic crystal for photonic and telecommunication applications. It has a broad optical transparency region spanning from the visible to the mid-infrared region, high refractive index \cite{Zelmon:97}, and favorable electro-optic properties \cite{Weis:1985, Garcia, Wooten}. These properties make LN an excellent candidate for integrated photonics \cite{Zhu:21, Boes:2018, Poberaj:2012}. The advancements in manufacturing wafer-scale, high-quality thin films of lithium niobate-on-insulator and breakthroughs in nanofabrication techniques have made LN nanophotonic components possible and capable of outperforming traditional photonic components\cite{Zhu:21}. In this paper, a lithium niobate photonic crystal has been used to design the valley PTI. While numerous works on silicon-based PTI have been done, lithium niobate has not yet been extensively demonstrated in this field, particularly in the telecommunication region. There has been an investigation on the rotated Weyl physics on nonlinear Lithium Niobate-on-Insulator Chips \cite{Yan} and a report of lithium niobate valley photonic crystal waveguide \cite{Rui}. However, in previous literature, there has been no demonstration of a valley photonic topological insulator on a lithium niobate platform that operates in the telecommunication region, which can also be tuned to a wide range of operating wavelengths with equal bandwidth in each case. Furthermore, previous works do not include demultiplexing operations and piezoelectric switching capabilities. The use of lithium niobate instead of the traditional silicon is motivated by two factors. The first factor is the wide transparency region of the LN material. The primary regions of operation shown in this work are the telecommunication regions of 1310 nm and 1550 nm. Using silicon for the PTI design with other operating wavelengths is difficult due to the high attenuation. However, the wide transparency region of lithium niobate allows the operating wavelength to be adjusted over a much more comprehensive range without significant attenuation. The second reason for using LN in our work is its prominent inverse piezoelectric effect \cite{Bouchy, Pop, Yao, Simeoni}. The operation of valley PTIs strongly depends on the spatial arrangement and the lattice dimensions. An external voltage induces stress in the lattice structures due to the inverse piezoelectric effect of LN. We show that if the magnitude of the external voltage is above a certain threshold, the stress-induced distortions in the lattice can remove topological protection at the operating wavelength. Based on this principle, we propose a piezoelectric switch-off mechanism for the optical channel.

Wavelength division demultiplexer is an essential component of optical communication systems and photonic integrated circuits \cite{Piggott:2015}. Its function is to isolate signals of target wavelengths from a multi-wavelength source into separate channels. Multiplexers and demultiplexers allow large transmission bandwidth simultaneously for multiple independent signals with high data throughput. Traditionally, demultiplexers are realized using arrayed waveguide gratings \cite{Dai:11}, echelle gratings \cite{Janz:2004}, and ring resonator arrays \cite{Xia:07}. But these devices are relatively large. Silicon photonic crystal-based implementations of demultiplexers \cite{Ooka:17, Bernier:08} have also been proposed with smaller footprints. This paper introduces a PTI-based demultiplexer design. The wavelength-selective nature of the edge modes and bulk insulation make PTIs ideal candidates for the design of integrated demultiplexers. The immunity to backscattering, sharp turns, and defects provide the added benefits of robust, low-loss transport through the channels. The tunability of the band gap and the broad transparency region of LN offer a wide range of operating wavelengths. Furthermore, the inverse piezoelectric effect of LN provides an electrical switch-off mechanism for the optical channels.

This paper proposes a PTI based on a lithium niobate photonic crystal and presents its operation as a 2-channel optical demultiplexer. The presence of the Dirac cone and band gap, the dependency of the band gap on the lattice structure dimensions, the transmittance spectrum of the channel, immunity to backscattering, sharp turns, and defects, and the tunability of the operating wavelength are investigated. The ability of the demultiplexer structure to isolate optical components into separate channels with crosstalk as low as $-$54 dB is verified. Lastly, the generation of lattice stress by an applied voltage and the resulting piezoelectric switch-off mechanism are presented. The authors envision that this work will expand the lithium niobate integrated photonics field and impact the development of efficient optical communication and photonic integrated circuits with tunable operating wavelengths and electrical switch-off mechanisms.     

\section{Methods}\label{4}

\subsection{Simulation setup}\label{4_1}
All optical simulations were done by solving Maxwell's equations using the finite-difference time-domain (FDTD) method \cite{Taflove}. The FDTD simulations were performed using the commercial software Lumerical FDTD: 3D Electromagnetic Simulator \cite{Lumerical}. The first step was generating the structure in Fig. \ref{fig: Fig. 1}(a). We wrote our custom code using the Lumerical scripting language to develop it. Each unit of the lattice is shown in Fig. \ref{fig: Fig. 1}(a). The dimensions of the structure depend on the lattice constant, $a$. The length of the equilateral triangles is $d_1 = d_2 = 0.5a$ for the photonic crystal lattice with a $C_6$ symmetry, and $d_1 = 0.55a$ and $d_2 = 0.45a$ for the photonic crystal lattice with a $C_3$ symmetry, the spacing between the triangles is 0.4$a$, and the width of the channel is also 0.4$a$. The structure consists of 82 units in the x direction and 40 units in the y direction, 20 units for each of the upper and lower photonic crystals. The dispersive refractive indices of the LN slab are shown in Fig. \ref{fig: Fig. 1}(b) \cite{Zelmon:97}. A TE mode source was used as input. Simulations were done with a mesh override region surrounding each triangle for accurate convergence. This is necessary for high resolution around the sharp corners and edges of the triangles. The maximum mesh step for the mesh override regions was chosen based on the smallest wavelength and the smallest dimension of the simulation. A maximum mesh step of 24 nm in the x, y, and z directions was taken for structures with $a>$ 850 nm, while a smaller mesh step of 6 nm was taken for structures with $a\leq$ 850 nm because of the smaller triangles. A separate scripting code generated the structure with the sharp turns. Defects were introduced by manually removing air holes near the channel. All post-processing was done in MATLAB.

\subsection{Band structure and valley Chern Number}\label{4_2}
The band structure was calculated using the commercial software Lumerical FDTD: 3D Electromagnetic Simulator \cite{Lumerical}. Randomly placed broadband dipole sources were used within an appropriate simulation region over our structure. The Bloch boundary conditions specified the wave vector k, meaning one simulation per k-vector is required. The electric fields will propagate indefinitely at frequencies where a mode or the band exists. The fields will quickly disappear at all other frequencies due to destructive interference. The bands were identified by finding the resonant frequencies of the fields that persist in the simulation. Using the field distribution and a discretized 2D Brillouin Zone, the valley Chern number was calculated by well-known methods in literature \cite{9445781, Wang:2020}. Calculations were done in MATLAB.

\subsection{Transmittance spectrum, spatial transmittance profile and normalized spatial power distribution} \label{4_3}
The source was placed at the start of the channel. A 2-dimensional (2D) frequency domain power monitor along the YZ plane was used to calculate the spatial transmittance profile in Fig. \ref{fig: Fig. 1}(c). A 1-dimensional (1D) frequency domain power monitor placed at the end of the channel with vertical dimensions equal to the channel width was used to calculate the transmittance spectrum and the spatial transmittance profile of the channel. A similar monitor placed at the start of the channel was used to calculate the backscattering from the channel. A 2-dimensional (2D) frequency domain power monitor was placed across the XY plane across the structure to calculate the normalized spatial power distributions. To find the relation between peak wavelength and lattice constant, we generated structures by sweeping the lattice constant from 600 nm to 1000 nm and analyzed the peaks from the transmittance spectrum in each case. Both 2D and 3D FDTD simulations were required.

\subsection{Piezoelectric simulations}\label{4_4}
The simulation of the inverse piezoelectric effect of LN was done using the finite element method (FEM) \cite{zienkiewicz}. At first, the entire structure was generated in the simulation environment. Ground was set to the lower boundary, and the input potential was set to the upper boundary to generate the potential distribution. A step potential was used as input. The strain-charge relation of the piezoelectric material, along with transient simulations, was done to generate the spatial stress profile and the magnitude of stress as a function of time. The necessary equations are shown in Section \ref{Section 3.3}. The transmittance spectrum and the normalized spatial power distributions of the stress-induced structure were then calculated using FDTD simulations.

\section{Results and Discussion}\label{sec2}
\subsection{Multi-wavelength photonic topological insulator}

\begin{figure}[t!]
\centering\includegraphics[width=1\textwidth]{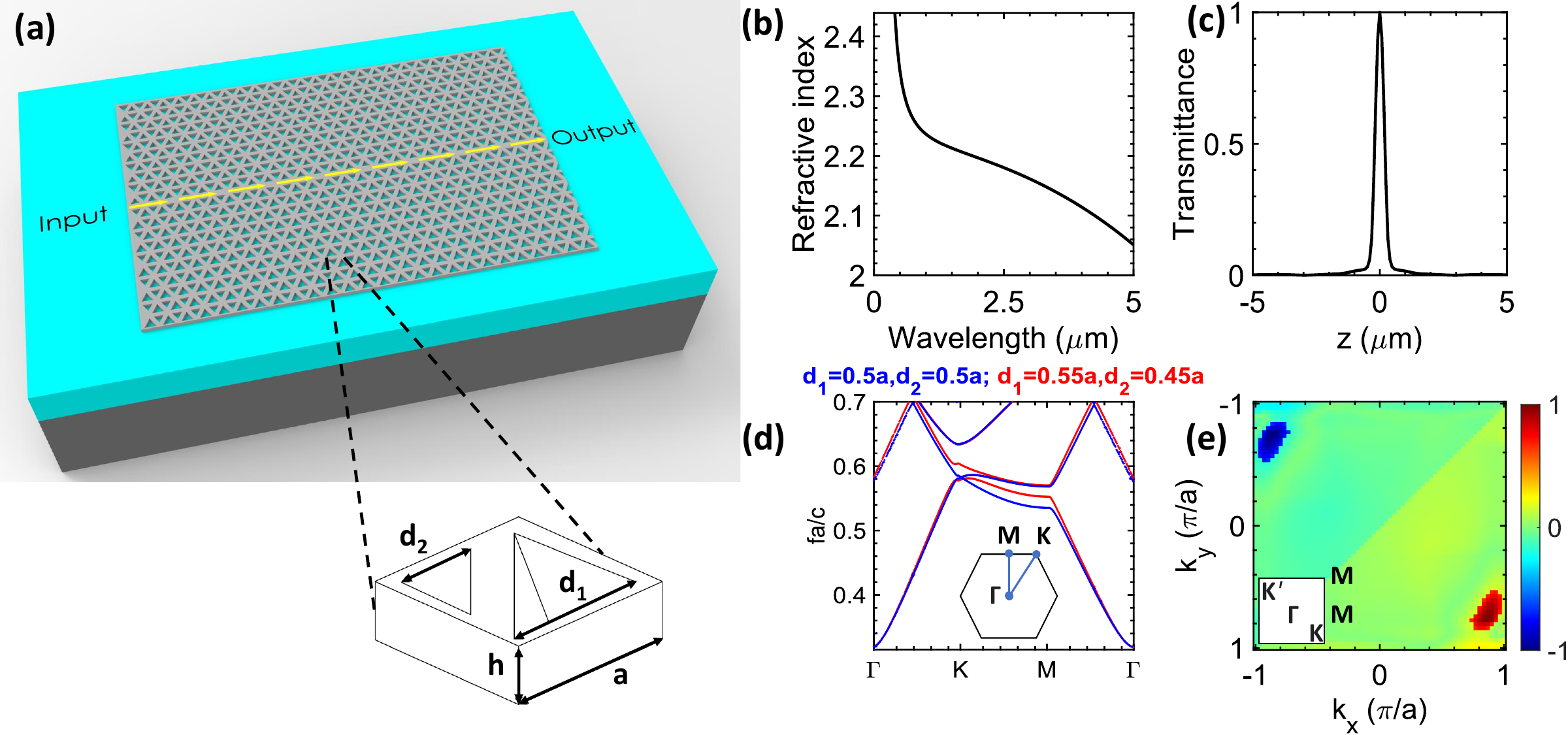}
\caption{Layout, photonic band structure, and Berry curvature. (\textbf{a}) Proposed layout of the structure. Two photonic crystals with honeycomb lattices formed on a lithium niobate slab with triangular air holes and an interface between the structures. Arrows indicate the predicted pathway of light through the interface between the two photonic crystals. A unit cell of the photonic crystal slab with the dimensions is shown. (\textbf{b}) The dispersive refractive index of the lithium niobate photonic crystal \cite{Zelmon:97}. (\textbf{c}) The spatial transmittance profile as a function of the z-axis position. The light remains confined to the slab instead of scattering into the air. (\textbf{d}) Photonic band structure for both the symmetric and asymmetric structures. $\Gamma$, $K$, and $M$ are the high symmetry points in the momentum space. The blue plot is for $d_1 = d_2 = 0.5a$, showing the Dirac cone at the $K$ point. The red plot is when spatial asymmetry is introduced by setting $d_1 = 0.55a$ and $d_2 = 0.45a$, which opens up the Dirac cone and introduces a band gap. The inset shows the Brillouin Zone path. (\textbf{e}) The normalized Berry curvature of the first band of the $C_3$-symmetric photonic crystal lattice in the momentum space spanned by $k_x$ and $k_y$, showing spikes at the $K$ and $K'$ points. The inset shows the Brillouin zone diagram of the hexagonal lattice.}\label{fig: Fig. 1}
\end{figure}

Our main design is shown in Fig. \ref{fig: Fig. 1}(a) with the predicted pathway of light. It consists of a lithium niobate slab with a honeycomb lattice. Each unit cell has a length equal to the lattice constant, $a$, and contains two inverted equilateral triangles, with lengths $d_1$ and $d_2$, as shown in Fig. \ref{fig: Fig. 1}(a). The periodic arrangement of these lattices in the slab forms two different photonic crystal structures with opposite spatial orientations. The positions of the large and small triangles are interchanged in the upper and lower photonic crystals. This forms a valley PTI with the interface between the two structures acting as the unidirectional pathway. The width of the channel is taken as 0.4$a$ to allow proper light propagation. The dispersive refractive index of the LN slab is shown in Fig. \ref{fig: Fig. 1}(b) \cite{Zelmon:97}. Light propagates only in the XY plane. It is confined in the z direction by total internal reflection, as shown by the spatial transmittance profile in Fig. \ref{fig: Fig. 1}(c).
\begin{figure}[t!]
\centering\includegraphics[width=1\textwidth]{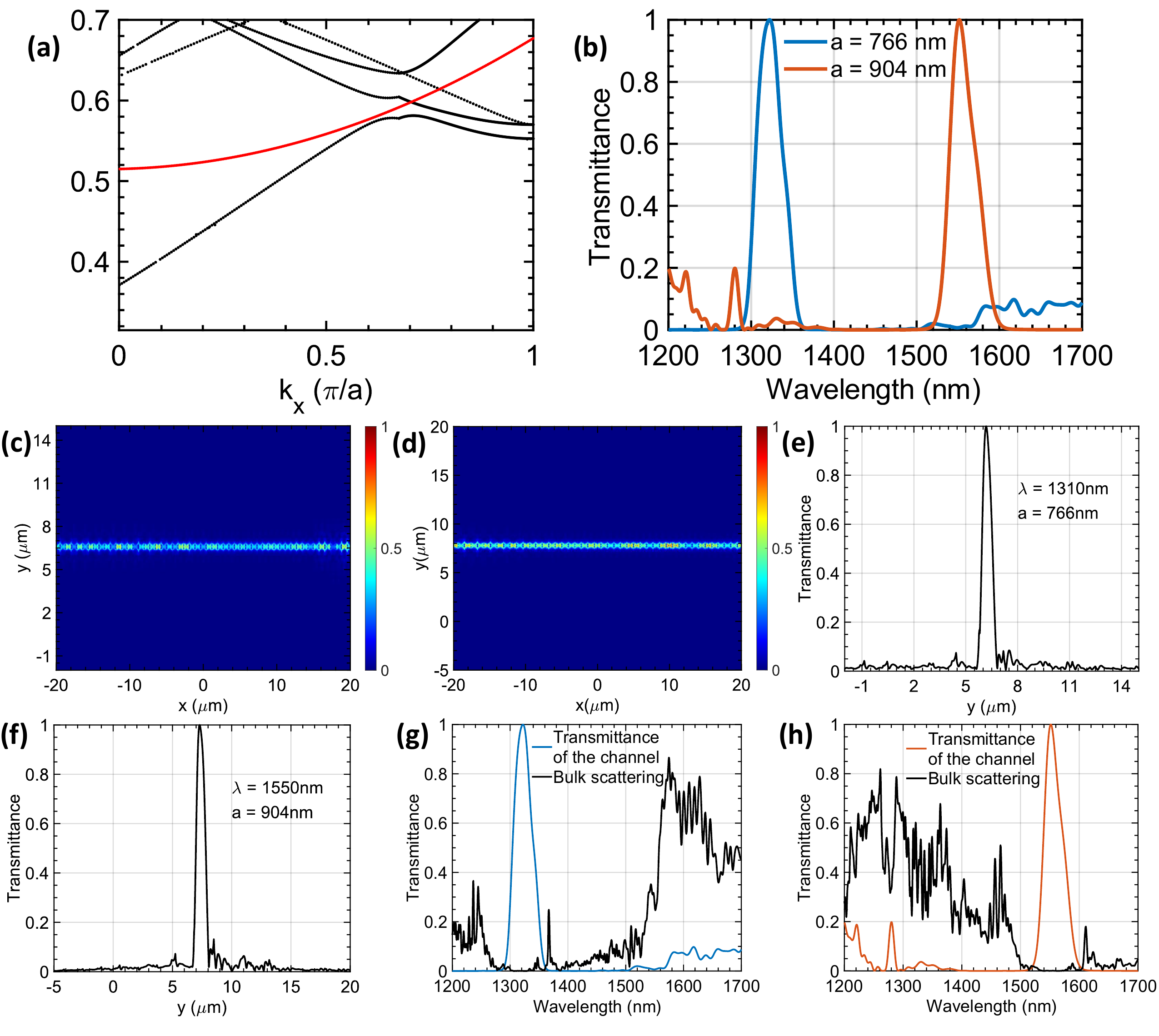}
\caption{Topologically protected propagation at the interface. (\textbf{a}) The projected band structure showing the edge state (red line). (\textbf{b}) Transmittance spectra of the channel for the structures with $a$ = 766 nm and $a$ = 904 nm. (\textbf{c}), (\textbf{d}) The normalized spatial power distributions across the XY plane of the entire structure. (\textbf{c}) is for the structure with $a$ = 766 nm and the normalized spatial power distribution corresponding to the wavelength, $\lambda$ = 1310 nm. (\textbf{d}) is for the structure with $a$ = 904 nm and the normalized spatial power distribution corresponding to the wavelength, $\lambda$ = 1550 nm. (\textbf{e}), (\textbf{f}) The spatial transmittance profiles as a function of the y-axis position taken from the end of the channel. (\textbf{e}) is for the structure with $a$ = 766 nm corresponding to the wavelength, $\lambda$ = 1310 nm. (\textbf{f}) is for the structure with $a$ = 904 nm corresponding to the wavelength, $\lambda$ = 1550 nm. (\textbf{g}) and (\textbf{h}) compare the transmittance along the channel and the spectrum scattered to the bulk. (\textbf{g}) is for the structure with $a$ = 766 nm, and (\textbf{h}) is for the structure with $a$ = 904 nm.}\label{fig: Fig. 2}
\end{figure}

Fig. \ref{fig: Fig. 1}(d) shows the photonic band structure of our design. The inset of Fig. \ref{fig: Fig. 1}(d) shows the Brillouin Zone path. We found a Dirac cone at the high symmetry $K$ point in the momentum space for the photonic crystal lattice with a $C_6$ symmetry, having $d_1 = d_2 = 0.5a$. Introducing asymmetry by setting $d_1 = 0.55a$ and $d_2 = 0.45a$ breaks the inversion symmetry of the lattice and reduces the lattice symmetry to $C_3$. This lifts the degeneracy at the $K$ point and opens up a band gap, as seen by the red plot in Fig. \ref{fig: Fig. 1}(d). Similar behaviors in the band structure were observed in previous works on Si PTI \cite{Shalaev:2019, Qi}. The parameters $d_1 = 0.55a$ and $d_2 = 0.45a$ were chosen because they open a well-defined band gap in the band structure at the desired location. The thickness of the slab, $h$, will depend on the availability of wafers. It is better to keep $h$ small to suppress higher-order slab modes in the out-of-plane direction. In Fig. \ref{fig: Fig. 1}(d), the frequency in the y-axis has been normalized. The location of the band gap and the Dirac cone are given in this normalized unit. The value of $a$ is adjusted based on the desired operating wavelength. The anisotropic nature of the LN crystal resulted in a directional band gap. This will cause imperfect insulation of the bulk. However, we present spatial transmittance profiles and the transmittance through the bulk that verify that the transmission is confined to the channel and the scattering into the bulk is negligible. 

Fig. \ref{fig: Fig. 1}(e) shows the Berry curvature calculated by the numerical method demonstrated in \cite{9445781}. The inset shows the Brillouin Zone diagram of the hexagonal lattice. Positive and negative spikes can be seen at $K$ and $K'$ points, respectively. The valley Chern numbers were calculated by integrating the Berry curvature over half of the first Brillouin Zone around the $K$ and $K'$ points:

\begin{equation}
    C^{K/K'} = \int_{HBZ} \Omega(k)d^{2}k \label{eq1}
\end{equation}
where $C^{K/K'}$ is the valley Chern number and $\Omega(k)$ is the Berry curvature as a function of $k$. Here, $C^{K/K'} \neq 0$. The valley Chern numbers for the upper and lower photonic crystals exhibit opposite signs. The topological edge state exists between the interface of the two crystals of opposite spatial `polarities,' where the orientation of the large triangle defines the polarity. The non-zero difference of the valley Chern numbers across the interface, $ | C_\nabla^{K/K'} - C_\Delta^{K/K'} | $, indicates the presence of edge states and the non-trivial topology of our design. The Dirac cone, band gap, Berry curvature, and valley Chern Number are the first pieces of evidence we present to justify our claim of topological protection. 

Through Fig. \ref{fig: Fig. 2}, we study the existence of the topologically protected edge state, wavelength selectivity, channel confinement, and bulk insulation of our design. Fig. \ref{fig: Fig. 2}(a) is the projected band structure showing the edge state (red line). The edge state is approximately linear in the band gap, indicating a constant group velocity. Our initial target wavelengths are the telecommunication regions of 1310 nm and 1550 nm. The structures for $a$ = 766 nm and $a$ = 904 nm were considered for achieving these operating wavelengths. The values of the lattice constant, $a$, are chosen so that the ratio $a$/$\lambda$ is at the center of the band gap. For these two particular cases, $a$/$\lambda$ = 766 nm/1310 nm = 904 nm/1550 nm $\approx$ 0.583.  Fig. \ref{fig: Fig. 2}(b) is the transmittance spectrum of the channel, calculated using the FDTD method. The vertical dimension of the monitor was equal to the width of the channel, so the spectrum shows only the components that propagate to the end of the channel. As we can see, high transmittance is observed around our desired wavelength while other components are suppressed. Similar shapes can be seen in the spectrum of previous PTI designs \cite{Shalaev:2019, Miguel}. The wavelength range of the topological edge states is localized around the operating wavelength. Spectral components outside the range of the edge mode cannot transmit through the channel. That is why we observe a bell-shaped transmittance with the spectrum diminishing as it moves further from the operating wavelength. Fig. \ref{fig: Fig. 2}(c) and \ref{fig: Fig. 2}(d) show the normalized spatial power distribution across the XY plane of the entire structure at $\lambda$ = 1310 nm and $\lambda$ = 1550 nm for $a$ = 766 nm and $a$ = 904 nm, respectively. In both cases, the optical signal is well confined within the channel and insulated from the bulk. No bulk scattering can be observed at the wavelengths of interest. To further verify that the operating wavelengths do not scatter into the bulk, we present the spatial transmittance profiles in Fig. \ref{fig: Fig. 2}(e) and \ref{fig: Fig. 2}(f). Each has been calculated with monitors at the end of the channel and plotted against the y-direction coordinates at operating wavelengths corresponding to their respective lattice constant. The operating wavelength remains confined to the channel. In Fig. \ref{fig: Fig. 2}(g) and \ref{fig: Fig. 2}(h), we compare the spectral components transmitted through the channel and the spectral components scattered into the bulk. Fig. \ref{fig: Fig. 2}(g) shows that the components around $\lambda$ = 1310 nm do not scatter into the bulk for the structure with $a$ = 766 nm, and Fig. \ref{fig: Fig. 2}(h) shows that the components around $\lambda$ = 1550 nm do not scatter into the bulk for the structure with $a$ = 904 nm. The absence of bulk scattering verifies the presence of the band gap at the operating wavelength, and the edge transmission of the operating wavelength quantitatively verifies the existence of an edge state.

\begin{figure}[t!]
\centering\includegraphics[width=0.95\textwidth]{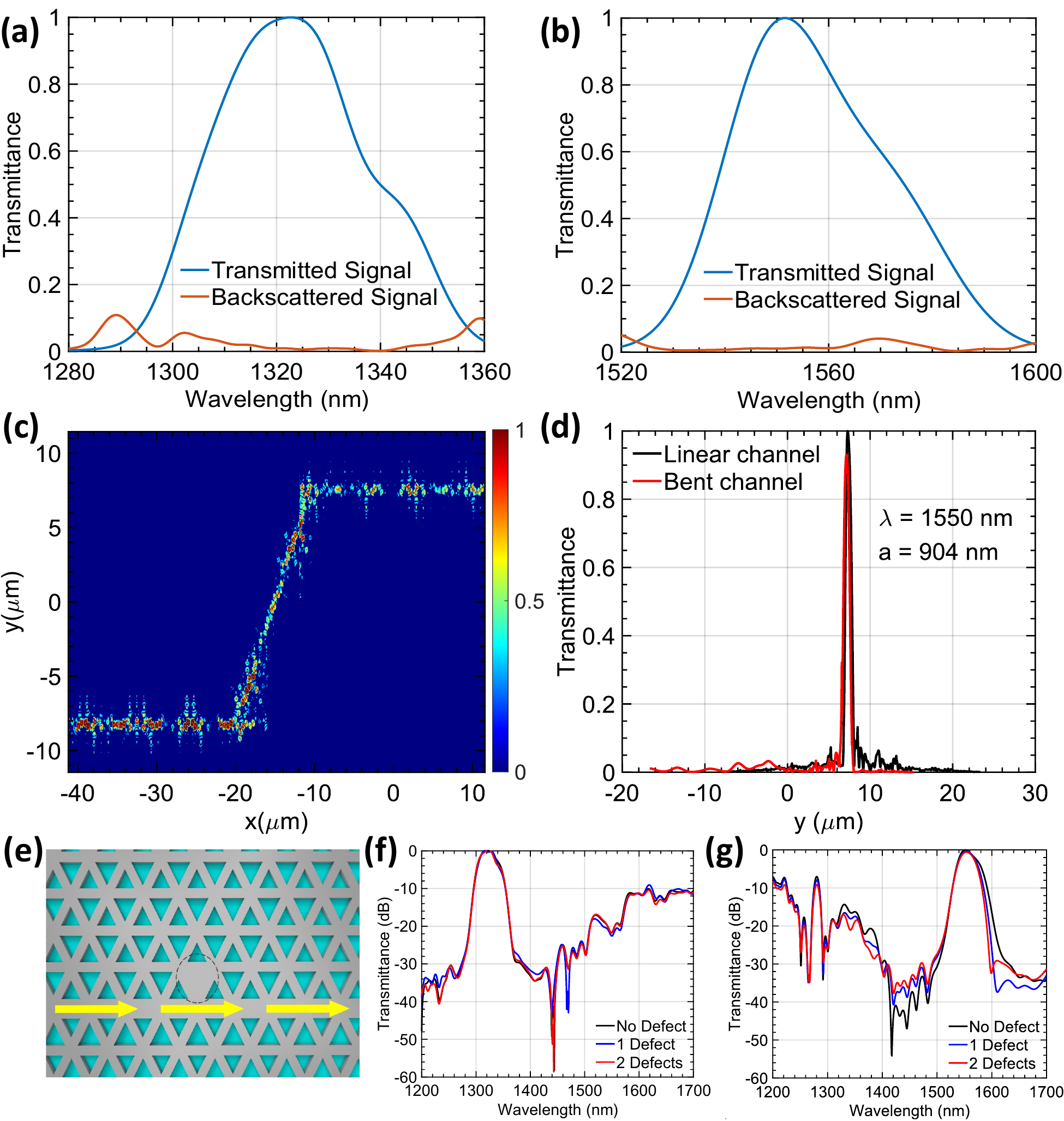}
\caption{Demonstration of robust transport. (\textbf{a}), (\textbf{b}) The transmittance spectra taken from the end of the channel and the backscattered signal taken from the start of the channel. (\textbf{a}) is for the structure with $a$ = 766 nm and (\textbf{b}) is for the structure with $a$ = 904 nm. (\textbf{c}) The normalized spatial power distribution across the XY plane of a larger structure with two sharp corners. (\textbf{d}) The spatial transmittance profile as a function of the y-axis position for the linear channel and the structure shown in (\textbf{c}). The profiles are taken from the end of the channel at $\lambda$ = 1550 nm. (\textbf{e}) The layout of the photonic crystal slab with a defect introduced at the corner of the transmission pathway in the form of a missing air gap. (\textbf{f}), (\textbf{g}) The transmittance spectra in dB for the case with no defects and after introducing 1 and 2 such defects. (\textbf{f}) is for the structure with $a$ = 766 nm and (\textbf{g}) is for the structure with $a$ = 904 nm.}\label{fig: Fig. 3}
\end{figure}

\begin{figure}[t!]
\centering\includegraphics[width=1\textwidth]{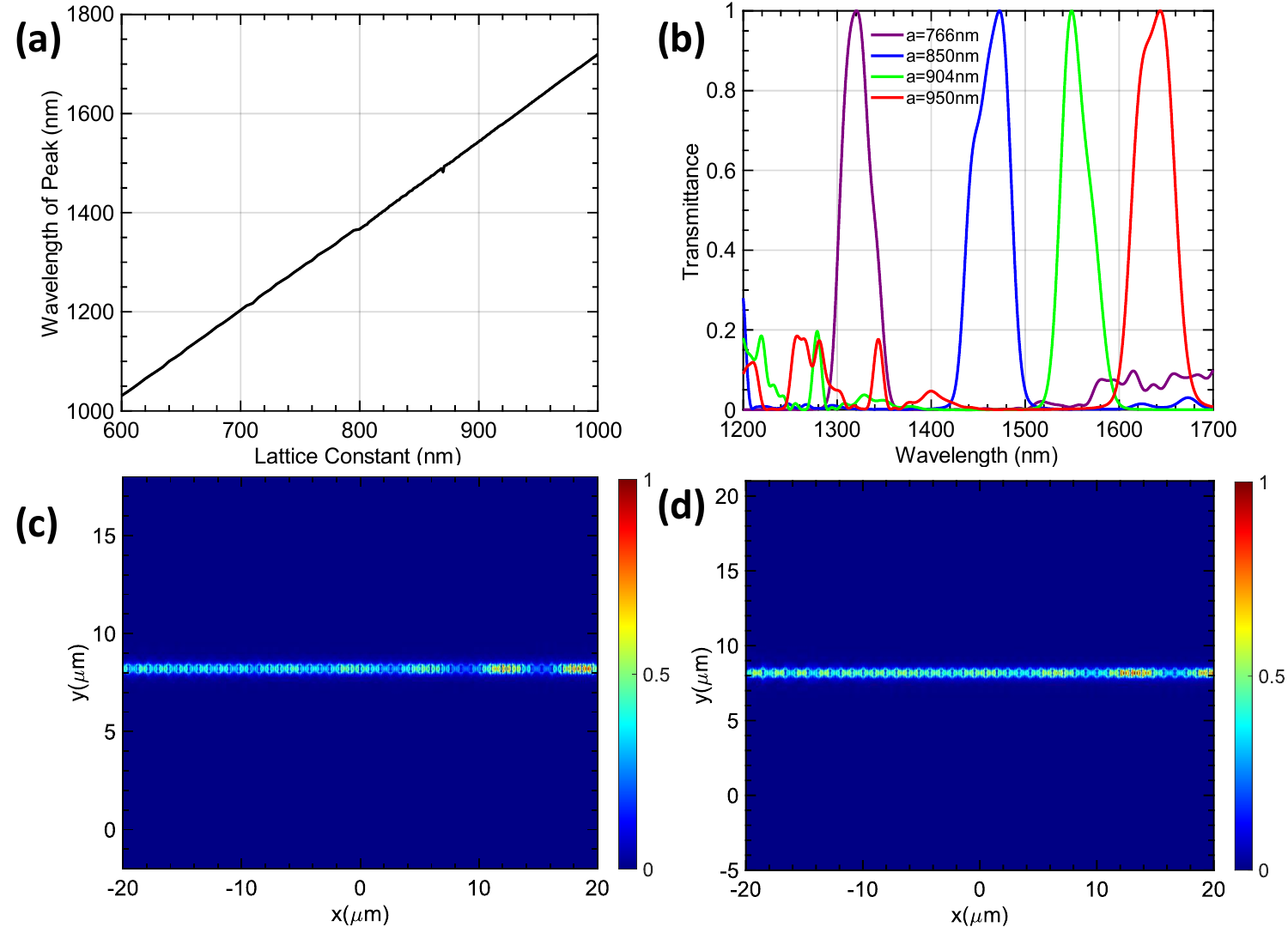}
\caption{Multi-wavelength operation. (\textbf{a}) The linear relationship between the lattice constant and the wavelength of the peak of the transmittance spectrum. (\textbf{b}) The transmittance spectra for different values of lattice constant. (\textbf{c}), (\textbf{d}) The normalized spatial power distributions across the XY plane of the entire structure. (\textbf{c}) is for the structure with $a$ = 850 nm and the normalized spatial power distribution corresponding to the wavelength, $\lambda$ = 1480 nm. (\textbf{d}) is for the structure with $a$ = 950 nm and the normalized spatial power distribution corresponding to the wavelength, $\lambda$ = 1644 nm.}\label{fig: Fig. 4}
\end{figure}
\begin{figure}[t!]
\centering\includegraphics[width=0.9\textwidth]{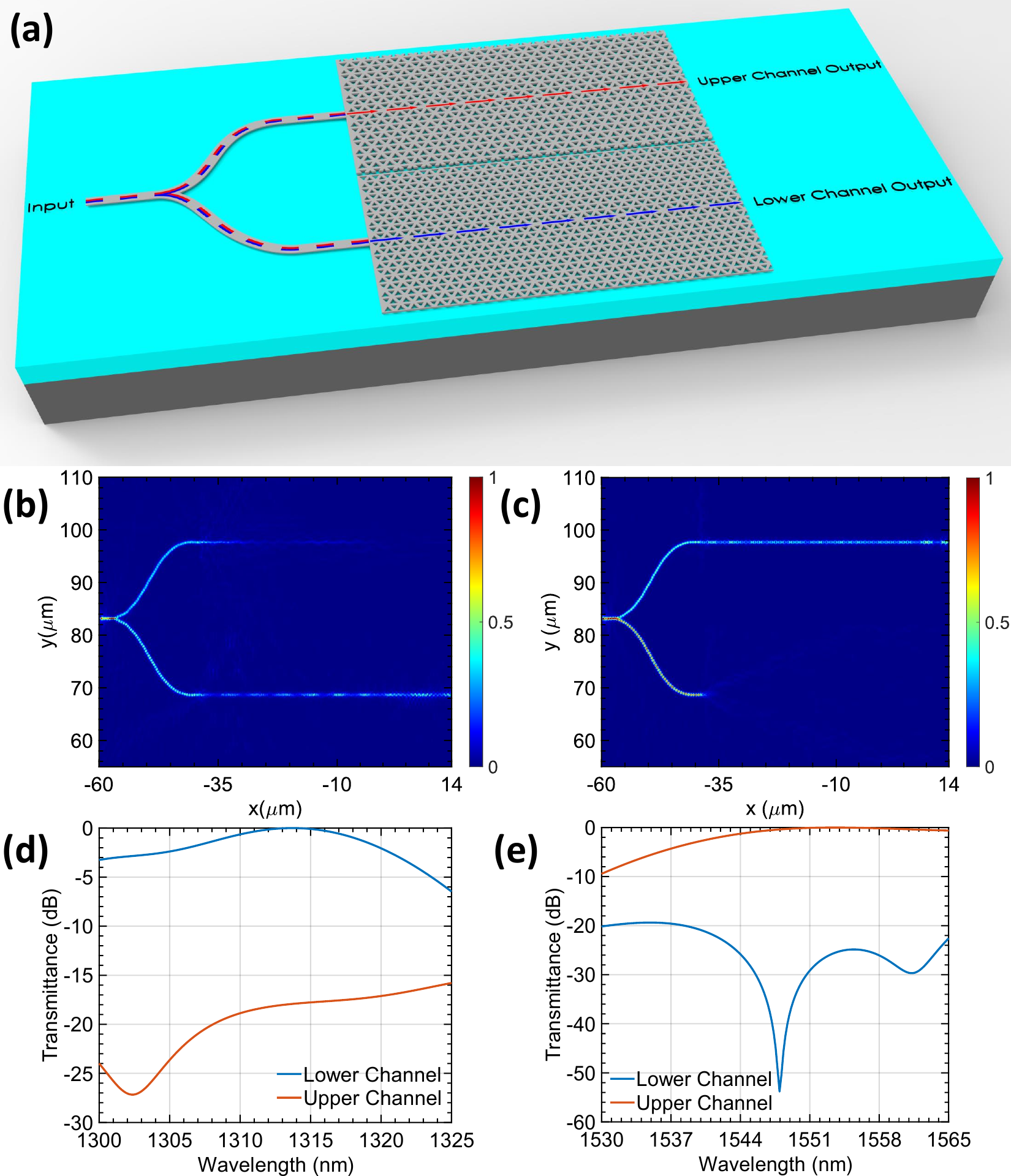}
\caption{Performance of the photonic topological insulator as an optical demultiplexer. (\textbf{a}) The layout of the demultiplexer structure. Two structures with $a$ = 766 nm and $a$ = 904 nm are stacked side by side. A lithium niobate waveguide is used to inject the input into the channels. The resulting structure has two channels and two output ports for the separated components. (\textbf{b}), (\textbf{c}) The normalized spatial power distribution across the XY plane of the demultiplexer structure. (\textbf{b}) is the normalized spatial power distribution corresponding to the wavelength, $\lambda$ = 1310 nm, and (\textbf{c}) is the normalized spatial power distribution corresponding to the wavelength, $\lambda$ = 1550 nm. (\textbf{d}), (\textbf{e}) The transmittance spectra in dB for both channels. (\textbf{d}) is for wavelengths from 1300 nm to 1325 nm, and (\textbf{e}) is for wavelengths from 1530 nm to 1565 nm.}\label{fig: Fig. 5}
\end{figure}
Fig. \ref{fig: Fig. 3} demonstrates robust transport. First, we study the defining characteristic of photonic topological insulators: backscattering immunity. The backscattered signal is calculated using a monitor at the start of the channel with a vertical dimension equal to the width of the channel. Fig. \ref{fig: Fig. 3}(a) and \ref{fig: Fig. 3}(b) show that almost no signal is backscattered in the operating wavelength regions. In Fig. \ref{fig: Fig. 3}(c), a larger structure with two sharp turns of the same photonic crystal design is considered. The angular segment of the channel is about 17 times the lattice constant. Despite this relatively large angular bend, the normalized spatial power distribution shows that the optical power is still confined within the channel despite the turns. Fig. \ref{fig: Fig. 3}(d) compares the spatial transmittance profiles at $\lambda$ = 1550 nm for the linear channel and the bent channel of the structure in Fig. \ref{fig: Fig. 3}(c). The spatial transmittance profiles have been observed to study the channel confinement of the optical signal despite the channel bend. The strength of transmittance for the bent channel is around 93$\%$ of the transmittance for the linear channel. The linear and bent channels have transmittance profiles confined around y = 7.2 $\mu$m at the operating wavelength of $\lambda$ = 1550 nm. The y-axis of Fig. \ref{fig: Fig. 2}(d) and Fig. \ref{fig: Fig. 3}(c) shows that this is the correct y-coordinate of the channel output port. The optical signal scattered into the bulk is less than 6$\%$ of the optical signal transmitted through the channel. Thus, our design is robust to sharp turns, and light of the operating wavelength is confined to the channel without bulk scattering. Next, we study the immunity to defects. Fabrication defects are a common issue in nanoscale structures. We introduce such a defect into our structure in the form of missing air holes, as shown in Fig. \ref{fig: Fig. 3}(e). The missing air holes are taken at a corner of the transmission pathway. For non-topological photonic crystal waveguides, a defect of this type near the channel would cause the optical signal to enter the bulk due to the interruption in the periodic lattice structure and the disruption of the periodic contrast of refractive indices. Fig. \ref{fig: Fig. 3}(f) and \ref{fig: Fig. 3}(g) compare the transmittance spectrum in dB for the case of no defect, one defect, and two defects. They show a negligible decrease in the transmittance near the operating wavelengths. There are no noticeable shifts in the wavelength of the peak either. Thus, our design is immune to backscattering, sharp turns, and defects concerning transmittance around the operating wavelength.

In Fig. \ref{fig: Fig. 4}, we study the shift in the wavelength of the peak of the transmittance spectra or the operating wavelength due to a change in lattice constant, $a$. Observing the transmittance spectrum after sweeping the lattice constant through a wide range shows the linear relationship in Fig. \ref{fig: Fig. 4}(a). The operating wavelength varies from $\lambda$ = 1031 nm to $\lambda$ = 1720 nm for a change of lattice constant from $a$ = 600 nm to $a$ = 1000 nm. The relation between the lattice constant and the operating wavelength can be modeled as a straight line because the location of the center of the band gap, when plotted in terms of normalized frequency, is approximately the same irrespective of the value of $a$ or $\lambda$. That is $a$/$\lambda \approx$ 0.583. So, a change in the lattice constant will cause a corresponding shift of the operating wavelength.

\begin{equation}
    a \approx 0.583\lambda_{operating} \label{eq2}
\end{equation}

Eq. \ref{eq2} gives the value of the lattice constant, $a$, for the operating wavelength. The other parameters depend on $a$, so they will change accordingly. To further substantiate this claim, we consider two structures for $a$ = 850 nm and $a$ = 950 nm and compare the transmittance with the previous results. Fig. \ref{fig: Fig. 4}(b) shows that the transmittance spectrum taken from the end of the channel is similar for the four cases, but the peaks have shifted. Fig. \ref{fig: Fig. 4}(c) and \ref{fig: Fig. 4}(d) are the normalized spatial power distribution across the structures for $a$ = 850 nm and $a$ = 950 nm, respectively, taken at their peak wavelength. They show that the optical power is confined to the channel and insulated at the bulk despite the change in the lattice dimensions. So, our design can operate as a photonic topological insulator for a wide range of wavelengths. The large transparency window, high refractive index, and low material loss of LN permit this multi-wavelength feature and open up the possibility of using our design at numerous operating wavelengths.
\subsection{Demonstration as an optical demultiplexer}

An important application of the wavelength-selective nature of the transmittance spectrum of our PTI design is to isolate wavelength components, while the bulk insulation is useful for channel isolation and, hence, low crosstalk. Fig. \ref{fig: Fig. 5} presents the performance of our design as a 2-channel optical demultiplexer. We demonstrate operation at 1310 nm and 1550 nm. Fig. \ref{fig: Fig. 5}(a) shows the demultiplexer structure. Two structures with $a$ = 766 nm and $a$ = 904 nm are generated and stacked side by side. The resulting structure has two channels or interfaces with two input and two output ports. A y-splitter waveguide made of LN is used to couple the optical source into the channel input ports. LN is chosen as the waveguide material to maximize mode matching and minimize coupling loss at the interface of the waveguide and the slab for higher coupling efficiency. Fig. \ref{fig: Fig. 5}(b) and \ref{fig: Fig. 5}(c) demonstrate the normalized spatial power distribution across the structure at $\lambda$ = 1310 nm and $\lambda$ = 1550 nm, respectively. The y-splitter transmits both components. But the lower channel supports only the $\lambda$ = 1310 nm component, and the upper channel supports only the $\lambda$ = 1550 nm component. The topological protection forces the optical power to be insulated at the bulk and, hence, the two channels to be isolated. Fig. \ref{fig: Fig. 5}(d) and \ref{fig: Fig. 5}(e) are the plots of the transmittance spectra in dB of both channels taken within selective regions. The transmittance spectra show that each channel has high transmittance around its operating wavelength with good bandwidth while suppressing unwanted components. The crosstalk in the lower channel goes as low as $-$54 dB, and in the upper channel, it goes as low as $-$27 dB. The wavelengths of 1310 nm and 1550 nm have been used as the carrier wavelengths. However, using lithium niobate for the PTI allows for a wider selection of PTI operating wavelengths, hence, the demultiplexer carrier wavelengths. A demultiplexer structure for any two of the operating wavelengths mapped in Fig. \ref{fig: Fig. 4}(a) can be formed using the same design. All we require is a change in the lattice constant.
\subsection{Piezoelectric switch-off} \label{Section 3.3}
\begin{figure}[t!]
\centering\includegraphics[width=0.7\textwidth]{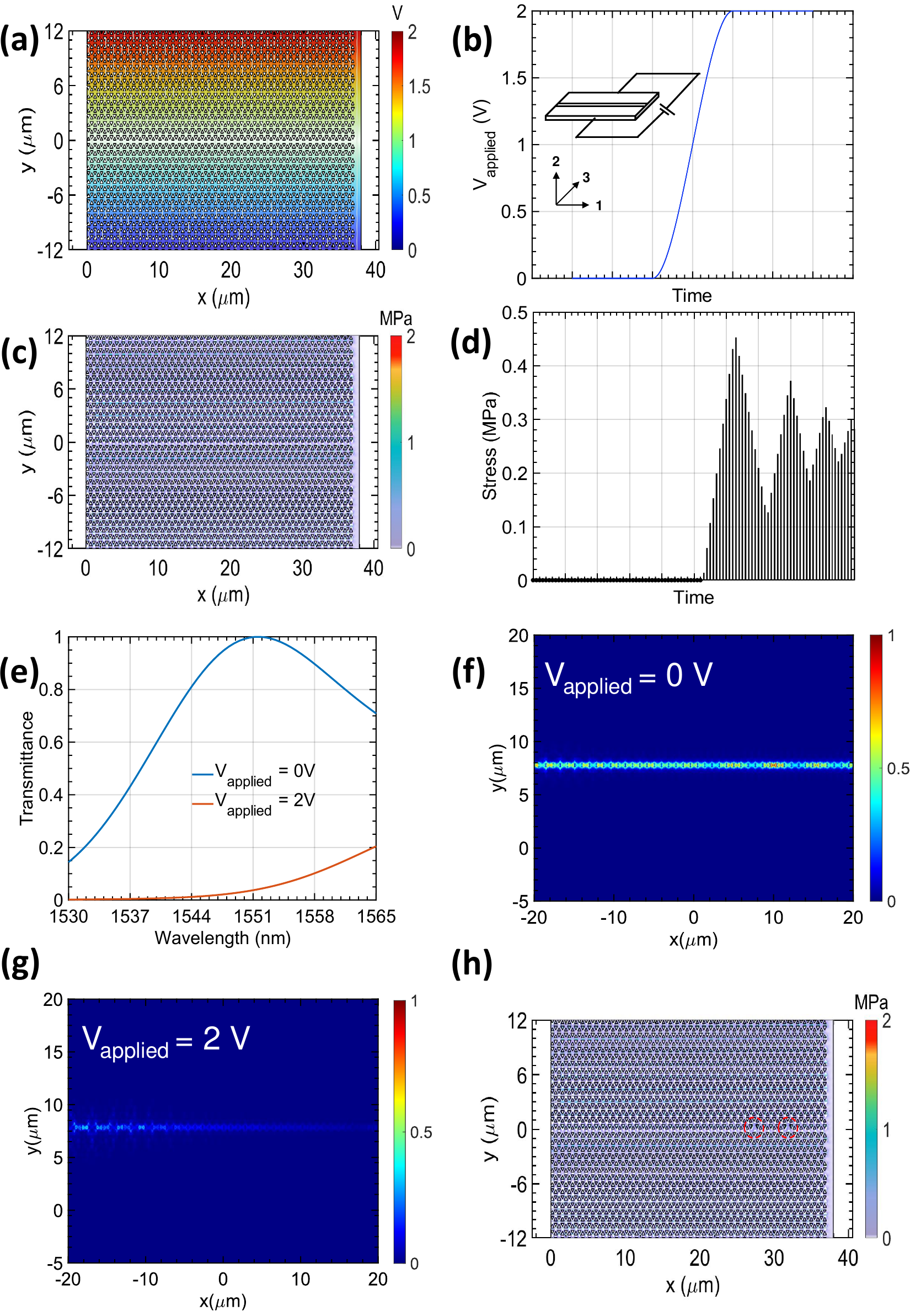}
\caption{Demonstration of piezoelectric switch-off. A potential difference has been applied across the structure with $a$ = 904 nm. (\textbf{a}) The spatial distribution of the applied voltage across the structure. (\textbf{b}) The magnitude of the applied step voltage as a function of time. The inset shows the setup and direction of the applied voltage on the LN slab. The "1", "2", and "3" axes in the inset are used to indicate the directions of the electric field and strain components. (\textbf{c}) The spatial profile of the induced stress across the structure and (\textbf{d}) the magnitude of that stress as a function of time. (\textbf{e}) The transmittance spectrum for the channel when the applied potential is respectively off and on. The wavelength range from 1530 nm to 1565 nm is shown. (\textbf{f}), (\textbf{g}) The normalized spatial power distribution across the XY plane of the structure corresponding to the operating wavelength. (\textbf{f}) is when the applied potential is off, and (\textbf{g}) is when the applied potential is on. (\textbf{h}) The spatial profile of the induced stress across the structure with defects. The dotted red circles indicate the defects.}\label{fig: Fig. 6}
\end{figure}
The topological protection and the transmission through the channel depend on the dimensions of the lattice structure, particularly the lattice constant and the length of the triangles. Sufficient mechanical stress on the lattice can strain the lattice dimensions. Such a strain on each lattice will affect the photonic band structure and, hence, topological protection and the transmittance of the operating wavelength. The electric field due to an external voltage will generate mechanical stress on the structure because of the inverse piezoelectric effect of LN. The stress-induced strain on each lattice affects channel transmission. We use FEM and FDTD simulations to investigate the effect of the external voltage on the transmission through the channel. The structure with $a$ = 904 nm is used for the study. We have simulated a potential difference using electrodes placed at the top and bottom of the plane of the LN slab, which contains the photonic crystal structure. Fig. \ref{fig: Fig. 6}(a) shows the spatial distribution of the applied potential across the structure. Fig. \ref{fig: Fig. 6}(b) shows the magnitude of the applied step voltage as a function of time. The inset of Fig. \ref{fig: Fig. 6}(b) shows the setup and direction of the electric potential. The axis "3" is along the applied electric field, the axis "1" is on the plane of the photonic crystal structure and perpendicular to the electric field, and the axis "2" is along the height of the slab. The electric field can be found from the voltage distribution, $\bm{E} = -\bm{\nabla} V$. The effect of the applied potential difference is studied using the strain-charge equation \cite{Ikeda}, given by
\begin{equation}
    \bm{S} = \bm{s_{E}T} + \bm{d^TE} \label{eq3}
\end{equation}
where $\bm{S}$, $\bm{T}$, and $\bm{E}$ are tensors that represent the induced strain, applied stress field, and applied electric field, respectively. $\bm{s_E}$ is the elastic compliance tensor at a constant electric field, and $\bm{d^T}$ is the transpose of the piezoelectric strain coefficient tensor. As shown in the inset of Fig. \ref{fig: Fig. 6}(b), the electric field is along axis "3", and no external stress field has been applied. Taking these factors into consideration and utilizing the crystal symmetry of LN, the strain components of interest on the plane of the photonic crystal structure can be expressed by
\begin{equation}
\begin{aligned}
    S_3 = d_{33}E_3 \\
    S_1 = d_{31}E_3 \label{eq4}
\end{aligned}  
\end{equation}
where $S_j$ is the strain component in the $j^{th}$ direction, $E_i$ is the electric field component in the $i_{th}$ direction, and $d_{ij} = (\frac{\partial S_j}{\partial E_i})_T$ is the piezoelectric strain coefficient at constant stress. The time dependency can be modeled with the governing equation \cite{Acoustic} written in terms of the structural displacement vector $\bm{u}$ and $V$ as given by
\begin{equation}
    \rho \frac{\partial^2\bm{u}}{\partial t^2} = \bm{\nabla}.(\bm{c_E}:\bm{\nabla u} + \bm{\nabla}V.\bm{e}) \label{eq5}
\end{equation}
where $\rho$ is the material density, $\bm{c_E}$ is the elastic constant at zero or constant electric field, and $\bm{e}$ is the piezoelectric stress coefficient. The induced strain is given by $\bm{S}=\frac{1}{2}[(\bm{\nabla u})^T+\bm{\nabla u}]$, and the induced stress is given by $\bm{T} = \bm{c_E S}$.

These equations model the inverse piezoelectric effect on the photonic crystal due to the applied voltage. The electric field due to the 2 V external voltage induces constant stress in the 1-3 plane of the structure, as shown in Fig. \ref{fig: Fig. 6}(c). Fig. \ref{fig: Fig. 6}(d) shows the magnitude of the stress as a function of time. The magnitude settles to 0.3 MPa. The inverse piezoelectric strain acts along the "1" and "3" axes, as given by the components in Eq. (\ref{eq4}). The boundary transmission of this strained structure is studied using FDTD simulations. The results are presented in Fig. \ref{fig: Fig. 6}(e), \ref{fig: Fig. 6}(f), and \ref{fig: Fig. 6}(g). The transmittance spectrum of the channel with the external voltage turned on and off is compared in Fig. \ref{fig: Fig. 6}(e). The wavelength range of $\lambda$ = 1530 nm to $\lambda$ = 1565 nm has been used considering the operating wavelength of the $a$ = 904 nm structure. When the applied voltage is turned off, the unstrained structure shows a transmittance peak at the operating wavelength. After a 2 V external voltage is applied, the transmittance curve of the strained structure shows that the operating wavelength decreases to less than 5$\%$ of its original value. A minimum of 2 V potential is required for the transmittance to be considered off. A higher voltage will further decrease the transmittance. However, a higher voltage will also induce a higher stress. A high stress across the structure might cause permanent defects in the photonic crystal. Considering these factors, the magnitude of the applied voltage and the generated stress are kept within typical reported values \cite{Pop, Yao, Simeoni}. Applying the voltage parallel to the plane of the photonic crystal structure instead of along the height of the slab allows strain components $S_1$ and $S_3$, parallel to the photonic crystal plane, to have higher values at acceptable voltages.  Fig. \ref{fig: Fig. 6}(f) and \ref{fig: Fig. 6}(g) show the normalized spatial power distribution across the XY plane of the structure at the operating wavelength. The results are for when the applied potential is respectively 0 V and 2 V. The high transmission through the channel and topological protection observed in the unstrained structure, as shown in Fig. \ref{fig: Fig. 6}(f), decreases significantly in the strained structure after the potential is applied, as shown in Fig. \ref{fig: Fig. 6}(g). Fig. \ref{fig: Fig. 6}(h) shows the effect of fabrication defects on the stress distribution. The defects are modeled as missing air gaps near the channel, indicated by the dotted red circles in the figure. According to Eq. (\ref{eq4}) and (\ref{eq5}), the inverse piezoelectric response depends on the voltage distribution and the electric field due to the distribution. If the number of defects does not significantly alter the voltage distribution, the stress distribution will remain the same. Fig. \ref{fig: Fig. 6}(g) demonstrates that introducing defects does not change the stress distribution, and the settled value remains 0.3 MPa. Thus, a minimum of 2 V applied potential removes topological protection and turns off transmission through the channel. Removing the applied voltage removes the induced strain and restores transmission. These results introduce a piezoelectric method to switch off the topological properties and transmission through the channel. Although the results for the structure with $a$ = 904 nm were presented, the mechanism also applies to structures with other lattice constants. Each channel of the demultiplexer can be similarly switched by this piezoelectric method.

\section{Conclusion}
In conclusion, a valley photonic topologically insulating 2-channel demultiplexer based on a lithium niobate photonic crystal with a wide range of operating wavelengths and a piezoelectric switch-off mechanism has been designed and demonstrated in this work. The existence of edge states, robust topological transport along the interface, optical insulation at the bulk, and immunity to backscattering, sharp turns, and defects was verified. A simple linear relation between the operating wavelength and the lattice constant of the structure was derived. Topological protection was preserved for each case of wavelength shift, thus introducing a simple method of designing the topological structure for multiple wavelength regions. An optical demultiplexer was formed using our design, which was shown to separate $\lambda$ = 1310 nm and $\lambda$ = 1550 nm components in isolated channels with good bandwidth. Topological insulation allows the crosstalk to be as low as $-$54 dB. Lastly, a method of piezoelectric switching of the topological protection was demonstrated, which can decrease the optical transmittance to less than 5$\%$ with a 2 V external potential. Other possible applications of our design include a piezoelectric optical filter, etc. This topological design will be helpful in applications that require lossless, robust optical transport, separation of wavelength components, and electrical switching of the optical transmittance. It will contribute to lithium niobate photonics, efficient optical communication systems, and photonic integrated circuits.



\begin{backmatter}
\bmsection{Funding}
Not applicable.

\bmsection{Acknowledgments}
Not applicable.

\bmsection{Disclosures}
\noindent The authors declare no conflicts of interest.

\bmsection{Data availability} Data underlying the results presented in this paper are not publicly available but may be obtained from the authors upon reasonable request.

\end{backmatter}

\bibliography{sample}






\end{document}